\documentclass[a4paper,11pt]{article}

\pdfoutput=1
\usepackage{jcappub}

\usepackage{bbm}
\usepackage{mathrsfs}
\usepackage{slashed}
\usepackage{caption}
\usepackage{epstopdf}
\usepackage[normalem]{ulem}
\usepackage[bottom]{footmisc}
\usepackage{subcaption}
\usepackage{bbold}
\usepackage{titlesec}
\usepackage{threeparttable}
\usepackage{booktabs}
\usepackage{changepage}
\usepackage[utf8]{inputenc}
\usepackage{dsfont} 
\usepackage{grffile}
\usepackage{graphicx}
\usepackage{dcolumn}
\usepackage{bm}
\usepackage{amssymb}
\usepackage{setspace}
\usepackage{amsmath}
\usepackage{array}
\usepackage{float}
\usepackage{lmodern}
\usepackage{multirow}
\usepackage{soul}
\usepackage{braket}
\usepackage{comment}
\usepackage[draft]{pgf}
\usepackage{adjustbox} 
\usepackage{xspace} 
\usepackage{url}
\usepackage{xcolor}
\usepackage{orcidlink}
\usepackage{chngcntr}
\usepackage{indentfirst}

\usepackage{physics}

\usepackage{mathtools,slashed}

\newcommand{\mi}{\mathrm{i}}
\newcommand{\me}{\mathrm{e}}

\allowdisplaybreaks

\newcommand{\be}{\begin{equation}}
\newcommand{\ee}{\end{equation}}

\newcommand{\bea}{\begin{eqnarray}}
\newcommand{\eea}{\end{eqnarray}}

\usepackage{orcidlink}

\begin{document}

\title{Cavity Multimodes as an Array for High-Frequency Gravitational Waves}

\author{
Yanfei Shang$^{a,b}$\,\footnote{These authors contributed equally to this work.}\,\orcidlink{0009-0003-1729-0809},
Yuxin Liu$^{c,d}$\,\footnotemark[1]\,\orcidlink{0000-0003-2296-4460},
Diego Blas$^{e,f}$\,\orcidlink{0000-0003-2646-0112},
Yifan Chen$^{g,h}$\,\footnote{Corresponding authors.}\,\orcidlink{0000-0002-2507-8272},
and Jing Shu$^{a,b,i}$\,\footnotemark[2] \,\orcidlink{0000-0001-6569-403X}
}

\affiliation[a]{School of Physics and State Key Laboratory of Nuclear Physics and Technology, Peking University, Beijing 100871, China}
\affiliation[b]{Center for High Energy Physics, Peking University, Beijing 100871, China}

\affiliation[c]{School of Physical Sciences, University of Chinese Academy of Sciences, Beijing 100049, China}
\affiliation[d]{International Centre for Theoretical Physics Asia-Pacific, University of Chinese Academy of Sciences (UCAS), Beijing 100190, China}

\affiliation[e]{Institut de Física d’Altes Energies (IFAE), The Barcelona Institute of Science and Technology, Campus UAB, 08193 Bellaterra (Barcelona), Spain}
\affiliation[f]{Instituci\'{o} Catalana de Recerca i Estudis Avan\c{c}ats (ICREA), Passeig Llu\'{i}s Companys 23, 08010 Barcelona, Spain}

\affiliation[g]{State Key Laboratory of Dark Matter Physics, Tsung-Dao Lee Institute, Shanghai Jiao Tong University, Shanghai 200240, China}
\affiliation[h]{Key Laboratory for Particle Astrophysics and Cosmology (MOE) \& Shanghai Key Laboratory for Particle Physics and Cosmology, Shanghai Jiao Tong University, Shanghai 200240, China}

\affiliation[i]{Beijing Laser Acceleration Innovation Center, Huairou, Beijing 101400, China}

\emailAdd{yanfeishang@stu.pku.edu.cn}
\emailAdd{liuyuxin211@mails.ucas.ac.cn}
\emailAdd{dblas@ifae.es}
\emailAdd{chen.yifan@sjtu.edu.cn}
\emailAdd{jshu@pku.edu.cn}

\abstract{
Microwave cavities operated in the presence of a background magnetic field provide a promising avenue for detecting high-frequency gravitational waves (HFGWs). We demonstrate for the first time that the distinct antenna patterns of multiple electromagnetic modes within a single cavity enable localization and reconstruction of key properties of an incoming HFGW signal, including its polarization ratio and frequency drift rate. Using a 9-cell cavity commonly employed in particle accelerators as a representative example, we analyze the time-domain response of 18 nearly degenerate modes, which can be sequentially excited by a frequency-drifting signal. The sensitivity is further enhanced by the number of available modes, in close analogy to the scaling achieved by a network of independent detectors, enabling sensitivity to astrophysically plausible binary sources.}

\date{\today}

\maketitle

\tableofcontents

\section{Introduction}

Gravitational wave (GW) observations across multiple frequency bands, from kHz signals of stellar-mass binaries detected by ground-based interferometers~\cite{LIGOScientific:2016aoc} to nHz backgrounds observed by pulsar timing arrays (PTAs), have opened new windows into the Universe. At frequencies far above the kHz range, no strong astrophysical sources are known, making the high-frequency GW (HFGW) band a promising arena for exploring new physics such as primordial black holes (PBHs) or bosonic clumps~\cite{Aggarwal:2020olq,Aggarwal:2025noe}. Detection in this regime remains largely uncharted. Several detector concepts first developed for axion searches, particularly electromagnetic (EM) resonant systems such as microwave and superconducting cavities, have recently been adapted for HFGW detection~\cite{Cruise:2000za,Li:2003tv,Ballantini:2003nt,Ballantini:2005am,Goryachev:2014yra,Herman:2020wao,Berlin:2021txa,Domcke:2022rgu,Berlin:2022hfx,Tobar:2022pie,Ahn:2023mrg,Alesini:2023qed,Berlin:2023grv,Bringmann:2023gba,Chen:2023ryb,Domcke:2023bat,Gao:2023gph,Gao:2023ggo,Navarro:2023eii,Schmieden:2023fzn,Domcke:2024mfu,Carney:2024zzk,Domcke:2024eti,Schneemann:2024qli,Valero:2024ncz,Capdevilla:2024cby,Fischer:2024msc,Garcia-Cely:2025mgu,Campbell:2025mks,Pappas:2025zld,Capdevilla:2025omb,Kharzeev:2025lyu,Kim:2025izt,Liang:2025vfv,Takai:2025cyy,Schenk:2025ria,Marconatoetal}.

As in laser interferometers and PTA observations, extracting the properties of a detected signal is crucial for confirming its quadrupolar GW nature~\cite{Chen:2024xzw} and identifying its origin. Although cosmological HFGWs are strongly constrained by Big Bang nucleosynthesis (BBN) bounds~\cite{Planck:2018vyg}, nearby coherent or transient sources, such as PBH binaries, could still produce detectable signals~\cite{Aggarwal:2020olq,Aggarwal:2025noe}. For such sources, resolving the propagation direction, polarization pattern, and waveform is essential. Similar to ultralight dark matter searches~\cite{Foster:2020fln,Chen:2021bdr,Jiang:2023jhl,Sulai:2023zqw,SHANHE:2024tpr,Jiang:2024boi,Gavilan-Martin:2024nlo,Reina-Valero:2025rul,Arza:2025kuh,Wilson:2025lhq}, a network of detectors could in principle provide this information~\cite{Schmieden:2023fzn,Schneemann:2024qli,Amaral:2026bef}, but practical challenges such as frequency overlap and phase synchronization remain, particularly for transient events.

In this work, we demonstrate the use of multiple EM modes within a single cavity as an effective detector array for HFGWs, focusing on the nearly degenerate modes of a multi-cell cavity. Such multi-cell geometries, widely used in accelerator technology to improve field uniformity and energy transfer efficiency, naturally support a set of near-degenerate resonant modes~\cite{Aune:2000gb,Belomestnykh:2006wd,Jeong:2020mtp}. An inspiraling primordial black hole binary can sequentially excite these modes as its GW frequency increases~\cite{Franciolini:2022htd,Barrau:2023kuv,Gatti:2024mde,Profumo:2024okx}. By comparing the relative signal strengths and phases among the modes, which requires at least $5$ loud modes, we show for the first time that a single cavity can localize and reconstruct key features of the GW waveform.

HFGWs in the MHz-to-GHz regime, for which no strong astrophysical sources are known, represent an unexplored frontier in fundamental physics~\cite{Aggarwal:2020olq,Aggarwal:2025noe} (see \cite{Casalderrey-Solana:2022rrn} for possible MHz signals in mergers of neutron stars). Given current sensitivities~\cite{Aggarwal:2020olq,Aggarwal:2025noe,TitoDAgnolo:2024res,Guo:2025cza}, the detection of a cosmological background at these frequencies is strongly constrained by the BBN bound on the effective number of relativistic degrees of freedom~\cite{Planck:2018vyg}. Nevertheless, several possible local sources within the Galaxy, such as PBH binaries or bosonic clumps, may generate signals detectable by upcoming precision experiments~\cite{Aggarwal:2020olq,Aggarwal:2025noe}.

\section{Cavity Response to High-Frequency Gravitational Waves}

Among the various proposed detection approaches, EM resonant systems, such as microwave and superconducting cavities, have attracted growing interest due to their conceptual similarity to axion dark matter searches. We focus on the direct coupling of GWs to EM fields and their conversion into photon signals in the presence of a background EM field through the inverse Gertsenshtein effect~\cite{Gertsenshtein:1962kfm}. This interaction originates from the minimal coupling between electromagnetism and gravity,
\begin{equation}
    S = \int \mathrm{d}^4 x\,\sqrt{-g}\,
    \left(
    -\frac{1}{4}\,
    g^{\mu\rho} g^{\nu\sigma}
    F_{\mu\nu} F_{\rho\sigma}
    \right),
\end{equation}
where $F_{\mu\nu}$ is the EM field-strength tensor. Under a weak gravitational perturbation, the spacetime metric is expanded as
\begin{equation}
    g_{\mu\nu}
    =
    \eta_{\mu\nu}+h_{\mu\nu}
    +\mathcal{O}(h^2),
\end{equation}
with $\eta_{\mu\nu}\equiv{\rm diag}(-1,1,1,1)$. 
In the presence of a background EM field $F_0^{\mu\nu}$, the interaction term relevant for GW--EM conversion can be written as~\cite{Berlin:2021txa}
\begin{equation}
    S \supset
    \int \mathrm{d}^4x\,
    j^\mu_{\rm eff} A_\mu,
\end{equation}
where the effective current induced by the GW is
\begin{equation}
    j^{\mu}_{\mathrm{eff}}
    =
    \partial_{\nu}
    \left(
    \frac{1}{2}\,h F_0^{\mu\nu}
    + h^{\nu}_{~\sigma}F_0^{\sigma\mu}
    - h^{\mu}_{~\sigma}F_0^{\sigma\nu}
    \right),
    \label{jeff}
\end{equation}
with $h \equiv \eta_{\mu\nu}h^{\mu\nu}$. 
The equation of motion for the EM field in vacuum is accordingly modified to
\begin{equation}
    (\nabla^2-\partial_t^2)\vec{E}
    =
    \partial_t \vec{j}_{\rm eff}
    + \vec{\nabla}\rho_{\rm eff},
    \label{eq:MaxwellMerged}
\end{equation}
where $j^\mu_{\rm eff}\equiv(\rho_{\rm eff},\vec{j}_{\rm eff})$.

The signal EM field of interest corresponds to resonant cavity modes, which receive contributions only from the effective three-current $\vec{j}_{\rm eff}$~\cite{hill2009electromagnetic}. 
We therefore expand the electric field inside the cavity in terms of normalized eigenmodes,
\begin{equation}
    \vec{E}(t,\vec{x})
    =
    \sum_a e_a(t)\,\vec{E}_a(\vec{x}),
\end{equation}
where the cavity eigenmodes satisfy
\begin{equation}
    \nabla^2\vec{E}_a+\omega_a^2\vec{E}_a=0,
    \qquad
    \int \mathrm{d}V\,
    \vec{E}_a\cdot\vec{E}_b^*
    =
    \delta_{ab},
    \label{eq:modeorth}
\end{equation}
with eigenfrequencies $\omega_a$. 

Projecting Eq.~(\ref{eq:MaxwellMerged}) onto a given cavity mode, we obtain the equation of motion for the mode amplitude $e_a(t)$~\cite{SHANHE:2023kxz,SHANHE:2024tpr},
\begin{equation}
    \ddot{e}_a(t)
    +
    2\gamma_a\dot{e}_a(t)
    +
    \omega_a^2 e_a(t)
    =
    -
    \int
    \partial_t\vec{j}_{\rm eff}
    \cdot
    \vec{E}_a^*
    \,\mathrm{d}V,
    \label{time equation1}
\end{equation}
where $\gamma_a \equiv \omega_a / (2Q_a^L)$ denotes the cavity dissipation rate, $Q_a^L$ is the loaded quality factor, and the integral is evaluated over the cavity volume.

We work in the transverse-traceless (TT) gauge~\cite{Gue:2026kga,Ratzinger:2024spd,Kim:2025izt,Schenk:2025ria} \footnote{We thank Jordan Gué and Tom Krokotsch for helpful discussions on the use of TT gauge for our signals.} and expand an incoming GW with momentum $\vec{k}$ in terms of the two polarization states,
\begin{equation}
    h^{ij}(\vec{k}\,)
    =
    \left(
    h_+\,\epsilon_+^{ij}(\hat{k})
    +
    h_\times\,\epsilon_\times^{ij}(\hat{k})
    \right)
    \me^{-\mi\vec{k}\cdot\vec{x}},
\end{equation}
where $h_{+/\times}$ denote the corresponding GW amplitudes. 
The polarization tensors are defined as
\begin{equation}
    \epsilon^+_{ij}
    =
    \hat{l}_i\hat{l}_j
    -
    \hat{m}_i\hat{m}_j,
    \qquad
    \epsilon^\times_{ij}
    =
    \hat{l}_i\hat{m}_j
    +
    \hat{m}_i\hat{l}_j,
\end{equation}
where $\hat{k}$, $\hat{l}$, and $\hat{m}$ form an orthonormal triad. In the cavity Cartesian frame $(x,y,z)$, expressed in terms of spherical coordinates, these unit vectors take the form
\begin{equation}
\begin{aligned}
    \hat{k} &=
    \left( \sin\theta \cos\phi,\,
           \sin\theta \sin\phi,\,
           \cos\theta \right)^{\mathrm{T}}, \\
    \hat{l} &=
    \left( \cos\theta \cos\phi,\,
           \cos\theta \sin\phi,\,
          -\sin\theta \right)^{\mathrm{T}}, \\
    \hat{m} &=
    \left( -\sin\phi,\,
            \cos\phi,\,
            0 \right)^{\mathrm{T}} .
\end{aligned}
\end{equation}

For simplicity, we consider a static magnetic background field of uniform strength $B_0$, while noting that the discussion can be readily generalized to oscillating EM backgrounds, such as secondary cavity modes enabling heterodyne conversion to probe frequencies far below the cavity resonance through either EM or mechanical coupling~\cite{Berlin:2023grv,Goryachev:2018vjt,Berlin:2019ahk,Berlin:2020vrk,Thomson:2021zvq,Berlin:2022hfx,Thomson:2023moc,Li:2025pyi,Fischer:2024msc,Marconatoetal}.

In the presence of the background magnetic field, the effective current induced by the GW can be decomposed as
\begin{equation}
    \vec{j}_{\mathrm{eff}} (\vec{k}\,)
    \equiv
    B_0 |\vec{k}|^2V^{1/3}
    \left(
    h_+ \hat{j}_+
    +
    h_{\times}\hat{j}_\times
    \right),
\end{equation}
where $\hat{j}_{+/\times}$ are dimensionless spatial profiles determined by the GW polarization and momentum~\cite{Berlin:2021txa}. 
The induced signal in a given cavity mode is proportional to the overlap between the effective current and the cavity electric-field distribution. This is characterized by the dimensionless overlap function, or antenna pattern,
\begin{equation}
    \eta^a_{+/\times}(\vec{k}\,)
    \equiv
    \frac{1}{V^{1/2}}
    \int_V
    \hat{j}_{+/\times}
    \cdot
    \vec{E}_a^*
    \,\mathrm{d}V,
    \label{eta_definition}
\end{equation}
which depends on both the GW propagation direction and frequency. In practice, the signal is confined to a narrow bandwidth around the cavity resonance, and we therefore evaluate the overlap functions at $|\vec{k}|\simeq\omega_a$.

With these definitions, Eq.~(\ref{time equation1}) can be rewritten as
\begin{equation}
    \ddot{e}_a(t)
    +
    2\gamma_a\dot{e}_a(t)
    +
    \omega_a^2 e_a(t)
    \simeq
    -\mi\omega_a^3V^{5/6}B_0
    \sum_{A=+,\times}
    h_A\,\eta_A^a.
    \label{eq:eommerged}
\end{equation}
The measured signal therefore corresponds to a scalar projection of the tensorial GW field, combining the two polarization components through the mode-dependent overlap functions. Recovering the full strain information requires multiple independent mode responses, achievable either through a network of cavities or by exploiting multimode structures within a single cavity.

\section{Multimode Structure and Antenna Patterns of the 9-Cell Cavity}

\subsection{Multimode Structure of the 9-Cell TESLA Cavity}

A resonant cavity supports a discrete set of electromagnetic modes characterized by mode indices, with frequency spacing determined by its geometry~\cite{hill2009electromagnetic,Navarro:2023eii,Navarro-Madrid:2025qtm}. 
A particularly important configuration is the multi-cell cavity, consisting of a chain of resonant cells coupled through irises. Analogous to a system of coupled oscillators, each eigenmode of a single-cell cavity splits into $N$ collective modes in an $N$-cell structure, whose eigenfrequencies remain nearly degenerate and lie close to that of the corresponding single-cell mode~\cite{Aune:2000gb,Belomestnykh:2006wd}.

Within this multiplet, the $a$-th collective mode ($a=1,\ldots,N$) exhibits relative electric-field amplitudes and phases across the cells such that the field in the $j$-th cell is proportional to $\sin[(j - 1/2)a\pi/N]$.
As $a$ increases, the field configuration evolves from the fully in-phase mode, where all cells oscillate coherently ($a=1$), to modes in which neighboring cells oscillate out of phase ($a=N$). 
The corresponding eigenfrequencies increase sequentially and approximately follow a cosine-like dispersion relation, with the frequency splitting determined by the coupling strength through the irises~\cite{Belomestnykh:2006wd}.

In this work, we adopt the well-established 9-cell elliptical TESLA cavity as a representative benchmark configuration~\cite{Aune:2000gb,Belomestnykh:2006wd}. 
This structure is widely used in superconducting accelerators, operating in the fundamental $\pi$ mode ($a=9$) at $1.3$~GHz. The cavity geometry corresponds to an effective volume of approximately $25$~L and can achieve intrinsic quality factors up to $Q_0\sim10^{11}$ when fabricated from superconducting niobium. In practice, however, operation in the presence of a strong background magnetic field is expected to significantly reduce the achievable quality factor.

We focus on the lowest transverse-electric multiplet TE$_{111}$, characterized by the mode indices $m=1$ (azimuthal), $n=1$ (radial), and $p=1$ (axial). 
For the 9-cell structure, the collective-mode frequencies range from approximately $1.51$~GHz at $a=1$ to $1.76$~GHz at $a=9$. 
The azimuthal index $m=1$ further introduces a twofold degeneracy, corresponding to electric-field components proportional to $\cos\phi$ and $\sin\phi$ around the cavity symmetry axis. We denote these two families by $\mathrm{TE}_{111+}$ and $\mathrm{TE}_{111-}$, respectively. 
In realistic cavity geometries, a small breaking of axial symmetry induces a slight frequency splitting between the two families, yielding a total of $18$ distinct modes.

\begin{figure}[h]
    \centering
    \includegraphics[width=\linewidth]{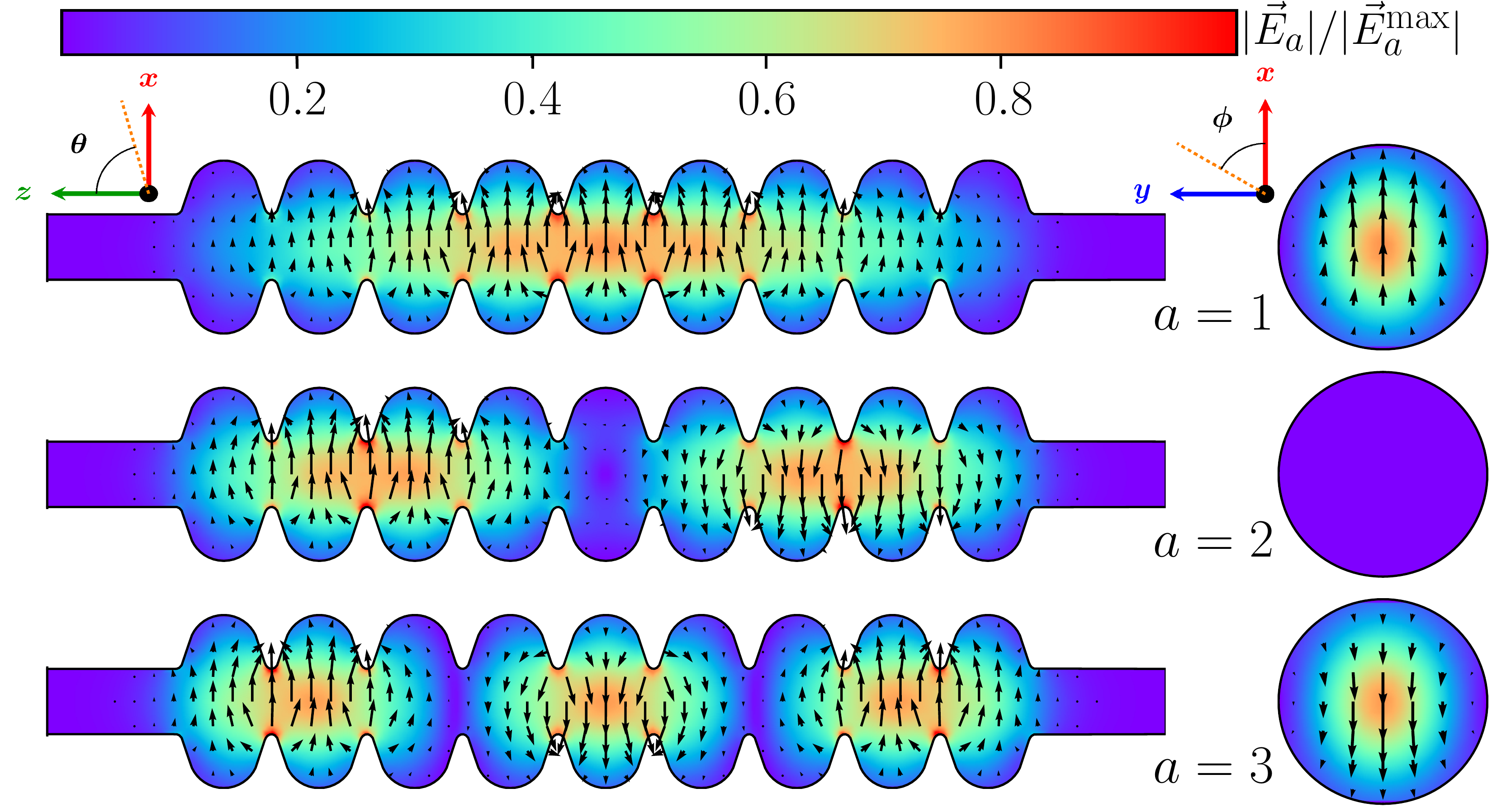}
    \caption{Electric-field configurations of the first three modes ($a = 1, 2, 3$) of the $\mathrm{TE}_{111+}$ family in a 9-cell elliptical TESLA cavity. The cavity symmetry axis is aligned with the $z$-axis, and the incoming direction is parameterized by the spherical coordinates $(\phi,\theta)$.  The left three panels show the electric fields on the $y=0$ plane and do not have perpendicular components. The right three panels show the fields on the $z=0$ plane in the center of the $5$-th cell, without perpendicular components. In each panel, arrow lengths are proportional to the normalized electric-field amplitude, normalized by the maximum value of each mode $|\vec{E}_a^{\rm max}|$. The corresponding $\mathrm{TE}_{111-}$ modes are obtained by a $\pi/2$ rotation about the cavity axis.}
    \label{fig:EMmode}
\end{figure}

Figure~\ref{fig:EMmode} shows the electric-field configurations of the lowest $\mathrm{TE}_{111+}$ collective modes projected onto the $y=0$ plane and onto the $z=0$ mid-plane of the fifth cell, with color indicating field magnitude and arrows showing field direction for the lowest collective modes ($a=1,2,3$).
The field amplitudes and relative phases across the cells clearly illustrate the multimode structure described above. The corresponding $\mathrm{TE}_{111-}$ modes are related by a rotation of $\pi/2$ about the cavity axis.

To lift the degeneracy between the $\mathrm{TE}_{111+}$ and $\mathrm{TE}_{111-}$ families, we introduce a small ellipticity of approximately $2\%$ into the cavity geometry. 
The mode properties are simulated using COMSOL, and the resulting resonant frequencies, loaded quality factors, coupling coefficients, overlap functions, and signal-to-noise-ratio contributions are summarized in Table~\ref{tab:fqletasnr}. To account for the degradation induced by the strong background magnetic field, we model the cavity walls using a fixed effective conductivity in the simulations. As a result, the intrinsic quality factors $Q_a^0$ vary slightly among different modes, with typical values of order $10^4$.

\begin{table*}[h]
    \centering
     \setlength{\tabcolsep}{7pt} 
\begin{tabular}{|c|c|c|c|c|c|c|c|c|c|c|}
\hline
\multicolumn{2}{|c|}{$a$} & 1 & 2 & 3 & 4 & 5 & 6 & 7 & 8 & 9 \\ \hline
\multirow{5}{*}[-0.4em]{$\mathrm{TE}_{111+}$} & $f_a$[GHz] & 1.51 & 1.52 & 1.54 & 1.56 & 1.60 & 1.63 & 1.67 & 1.71& 1.76 \\
\cline{2-11}
& $Q^L_a$ & 8600 & 3192 & 1640 & 1037 & 753 & 608 & 542 & 563 & 785 \\
\cline{2-11}
& $\beta_a$ & 1.1 & 4.7 & 10.4 & 17.6 & 26.2 & 34.4 & 40.6 & 40.1 & 33.2 \\
\cline{2-11}
& $|\eta_\mathrm{eff}^a|\times10^3$ & 1.0 & 12.4 & 20.4 & 13.1 & 0.6 & 5.1 & 1.3 & 2.7 & 2.6 \\
\cline{2-11}
& $\mathrm{SNR}^2_a$ & 0.44 & 24.36 & 34.89 & 9.63 & 0.02 & 0.96 & 0.06 & 0.30 & 0.41 \\
\hline
\multirow{5}{*}[-0.4em]{$\mathrm{TE}_{111-}$} & $f_a$[GHz] & 1.52 & 1.53 & 1.55 & 1.58 & 1.61 & 1.64 & 1.68 & 1.72 & 1.76 \\
\cline{2-11}
& $Q^L_a$ & 9112 & 3448& 1785 & 1136& 832 & 681 & 624 & 696 & 1557 \\
\cline{2-11}
& $\beta_a$ & 1.0 & 4.5 & 9.9 & 16.6 & 24.2 & 31.1 & 35.3 & 31.7 & 17.2 \\
\cline{2-11}
& $|\eta_\mathrm{eff}^a|\times10^3$ & 0.6 & 0.72 & 11.9 & 7.6 & 1.1 & 2.9 & 1.8 & 1.1 & 4.1 \\
\cline{2-11}
& $\mathrm{SNR}^2_a$ & 0.16 & 9.18 & 13.28 & 3.62 & 0.06 & 0.36 & 0.14 & 0.12 & 1.99 \\
\hline
\end{tabular}
\caption{Properties of the $18$ cavity modes, including the $\mathrm{TE}_{111+}$ and $\mathrm{TE}_{111-}$ families. Listed are the resonant frequencies $f_a\equiv\omega_a/(2\pi)$, loaded quality factors $Q_a^L$, and coupling coefficients $\beta_a$. Also shown are the effective overlap functions $|\eta_{\rm eff}^a|$ and the corresponding contributions to the squared signal-to-noise ratio, $\mathrm{SNR}_a^2$, for the benchmark GW signal.}
\label{tab:fqletasnr}
\end{table*}

\begin{figure}[h]
    \centering
    \includegraphics[width=\linewidth]{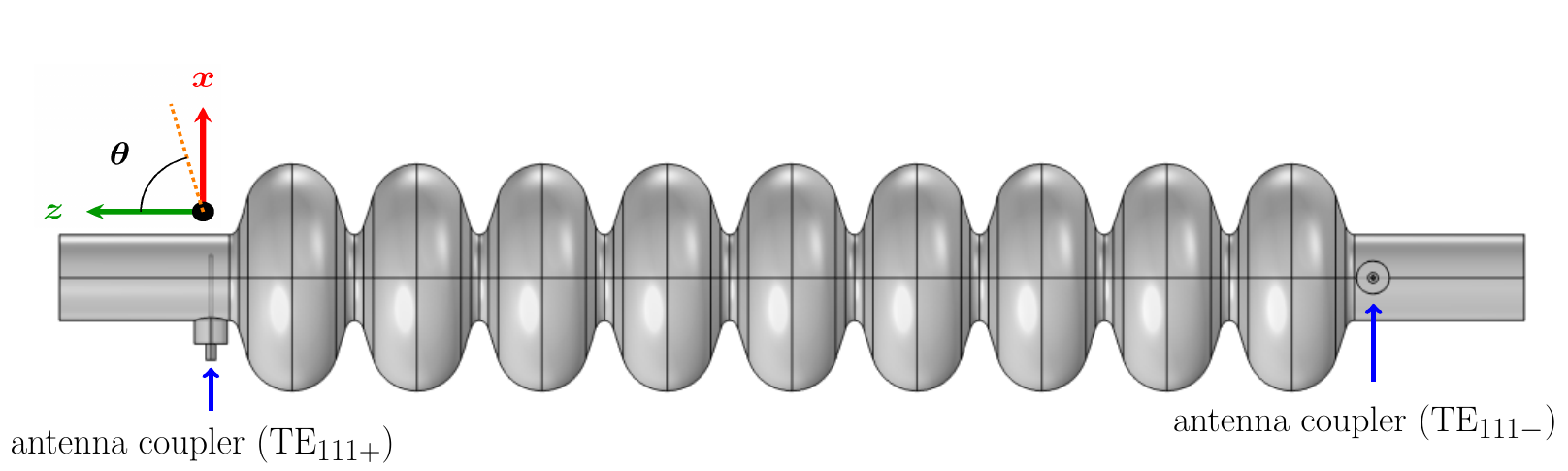}
    \caption{Side view of the 9-cell cavity along the $y$ direction. Two straight antenna couplers are employed: the left coupler predominantly reads out the $\mathrm{TE}_{111+}$ modes, while the right coupler reads out the $\mathrm{TE}_{111-}$ modes.}
    \label{fig:betaandantenna}
\end{figure}

The modes are read out using two straight antenna couplers oriented in orthogonal directions, as shown in Fig.~\ref{fig:betaandantenna}. The left antenna predominantly couples to the $\mathrm{TE}_{111+}$ family, while the right antenna couples to the $\mathrm{TE}_{111-}$ family. 
The corresponding coupling coefficients,
\begin{equation}
    \beta_a \equiv \frac{Q_a^0}{Q_a^L}-1,
\end{equation}
span the range from critical coupling to the over-coupled regime. The frequency separations between different modes can be readily covered by broadband amplifiers such as Josephson traveling-wave parametric amplifiers (JTWPAs)~\cite{ADMX:2021mio}.

\subsection{Overlap Functions and Antenna Patterns}

\begin{figure}[h]
    \centering
       \includegraphics[width=\linewidth]{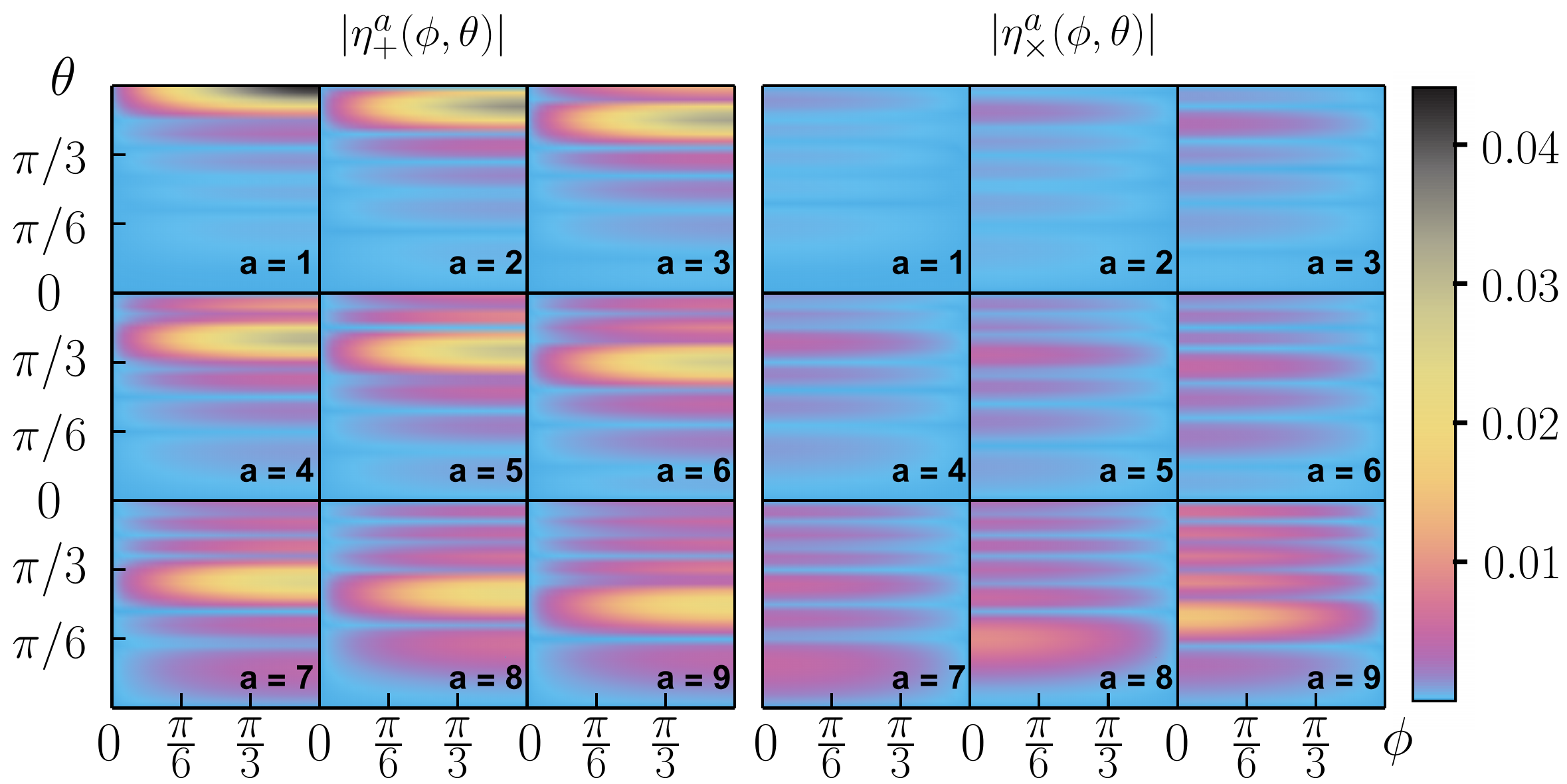}
    \caption{Overlap functions $\eta_{+,\times}^a$ for GWs incident from direction $\hat{k} \equiv (\phi, \theta)$, shown for the $\mathrm{TE}_{111+}$ modes of the 9-cell cavity. For each mode, the GW frequency is taken to match the corresponding cavity resonance $\omega_a$. The plots are shown for $\phi, \theta \in (0, \pi/2)$; overlap functions in other regions of the sky, as well as those for the $\mathrm{TE}_{111-}$ modes, can be obtained via the symmetry relations discussed in the text.}
    \label{fig:9cell_freqskymap}
\end{figure}

The overlap functions $\eta^a_{+/\times}$ are computed by extracting the electric-field configurations from COMSOL simulations and evaluating Eq.~(\ref{eta_definition}) at resonance, $|\vec{k}|\simeq\omega_a$, for each mode. 
Figure~\ref{fig:9cell_freqskymap} shows the resulting overlap functions for the $\mathrm{TE}_{111+}$ modes as functions of the GW propagation direction $\hat{k}\equiv(\phi,\theta)$, assuming a uniform magnetic field aligned with the $+z$ direction.

Each cavity mode exhibits a pronounced directional preference in the polar angle $\theta$. This behavior can be understood as a phase-matching condition between the GW plane-wave phase and the modulation of the electric-field configuration across the cavity cells. 
Denoting by $L_z=0.114~{\rm m}$ the distance between neighboring cell centers, the overlap functions approximately scale as
\begin{equation}
  \left|\eta^a_{+/\times}\right|
  \propto
  \left|
  \sum_{j=1}^{N}
  e^{ikjL_z\cos\theta}
  \sin\!\left[
  \left(j-\frac12\right)\frac{a\pi}{N}
  \right]
  \right|
  \propto
  \left|
  \frac{
  \sin\!\left(\frac{N}{2}\psi\right)
  }{
  \sin\!\left(\frac{\psi}{2}\right)
  }
  \right|,
\end{equation}
where
\begin{equation}
    \psi
    \equiv
    kL_z\cos\theta
    -
    \frac{a\pi}{N}.
    \label{eq:psi}
\end{equation}
The overlap therefore reaches its maximum when the phase-matching condition $\psi = 0$ is satisfied, explaining the sharp $\theta$ dependence observed in Fig.~\ref{fig:9cell_freqskymap}.

The azimuthal dependence arises from the projection of the effective current onto the dominant electric-field direction of the $\mathrm{TE}_{111+}$ modes. One finds
\begin{equation}
    \hat{j}_+\cdot\hat{x}
    \propto
    \sin\theta\,\sin\phi,
    \qquad
    \hat{j}_\times\cdot\hat{x}
    \propto
    \sin\theta\,\cos\theta\,\cos\phi,
\end{equation}
which reproduces the qualitative behavior observed in the numerical overlap functions. The additional factor of $\cos\theta$ further explains the relatively suppressed magnitude of $\eta_\times^a$ compared to $\eta_+^a$.

The overlap functions shown in Fig.~\ref{fig:9cell_freqskymap} are displayed only over one-eighth of the sky, corresponding to $\phi,\theta\in[0,\pi/2]$. 
The remaining regions can be obtained through discrete parity transformations acting on the combined system of the cavity, the background magnetic field, and the GW plane wave:
\begin{enumerate}
    \item[(i)]
    $\phi\rightarrow\pi+\phi$,
    corresponding to parity in the $x$--$y$ plane;
    \item[(ii)]
    $\phi\rightarrow\pi-\phi$,
    corresponding to parity along the $x$ direction;
    \item[(iii)]
    $\theta\rightarrow\pi-\theta$,
    corresponding to parity along the $z$ direction.
\end{enumerate}
Under these transformations, the overlap functions change only by overall signs determined by the parity properties of the GW polarization tensors, the background magnetic field, and the cavity eigenmodes. 
The resulting transformation properties for the $\mathrm{TE}_{111+}$ modes are summarized in Table~\ref{tab:eta_symmetry}.

\begin{table}[h]
    \centering
\begin{tabular}{|c|c|c|c|c|c|c|c|c|}
    \hline
     & $h_\times$ & $h_+$ & $F^{ij}_0$ & $\hat{j}_\times$ & $\hat{j}_+$ & $\vec{E}_a(\vec{r}\,)$ & $\eta_\times^a$ & $\eta_+^a$ \\
    \hline
    $\phi \rightarrow \pi + \phi$     &  1  & 1 & 1  & 1  & 1  & $-1$ & $-1$ & $-1$ \\
    \hline
    $\phi \rightarrow \pi - \phi$     & $-1$ & 1 & $-1$ & 1  & $-1$ & $-1$ & $-1$ & 1 \\
    \hline
    $\theta \rightarrow \pi - \theta$ & $-1$ & 1 & 1  & $-1$ & 1 & $(-1)^{a-1}$ & $(-1)^a$ & $(-1)^{a-1}$ \\
    \hline
\end{tabular}
\caption{Parity properties of the quantities entering the overlap functions $\eta_A^a$ for the $\mathrm{TE}_{111+}$ modes.}
\label{tab:eta_symmetry}
\end{table}

The $\mathrm{TE}_{111-}$ modes are related to the $\mathrm{TE}_{111+}$ family by a rotation of $\pi/2$ around the cavity axis, corresponding to the transformation $\phi\rightarrow\phi+\pi/2$.

These discrete symmetries imply that the overlap functions exhibit an apparent $Z_8$ degeneracy among different sky octants. 
When several cavity modes with different indices $a$ are measured simultaneously, however, their distinct symmetry properties, particularly under $\theta\rightarrow\pi-\theta$, allow partial reconstruction of the relative phase between the two GW polarizations. As a result, the degeneracy is reduced from eightfold to fourfold. The remaining degeneracy could in principle be lifted by introducing controlled inhomogeneities in the background magnetic field, although such configurations would require additional experimental complexity and calibration.

\section{Time-Domain Response to a Chirping HFGW Signal}

Multi-cell cavities support multiple nearly degenerate EM modes with distinct antenna patterns. The relative amplitudes and phases among these modes therefore provide a means to extract information about an incident GW signal.

\subsection{Benchmark Chirping GW Waveforms}

We consider a generic chirping HFGW signal whose instantaneous angular frequency evolves linearly in time,
\begin{equation}
    \omega_g(t) = \omega_1 + \alpha (t - t_0),
\end{equation}
where $t_0$ denotes the time at which the signal frequency crosses the first cavity resonance $\omega_1$, and $\alpha$ is the angular-frequency drift rate characterizing the temporal evolution of the signal. Such a chirping waveform can arise from compact binary inspirals, including PBH binaries, as a possible physical realization.

The two polarization components are modeled as~\cite{Maggiore:2018sht}
\begin{equation}
\begin{aligned}
h_{+}(t)
&=
\frac{h_0}{\sqrt{1+\kappa^2}}
\exp\left[
\mi\left(
\delta_0+\omega_1(t-t_0)+\frac{\alpha}{2}(t-t_0)^2
\right)
\right],\\
h_{\times}(t)
&=
\frac{h_0\kappa}{\sqrt{1+\kappa^2}}
\exp\left[
\mi\left(
\delta_0+\xi+\omega_1(t-t_0)+\frac{\alpha}{2}(t-t_0)^2
\right)
\right].
\end{aligned}
\label{eq:hPBH}
\end{equation}
Here $h_0$ is the overall strain amplitude, $\kappa$ is the polarization amplitude ratio, $\xi$ is the relative phase between the two polarizations, and $\delta_0$ is the initial phase.

A more accurate chirp model can be incorporated straightforwardly without introducing additional parameters for circular inspirals. More complex waveforms, including eccentric binaries or signals beyond the standard circular-orbit case~\cite{Barrau:2023kuv,Teuscher:2024xft,Barrau:2024kcb,Cusin:2024git,Jamet:2024jca}, are left for future work.

We consider a benchmark set of waveform parameters with identical angular, polarization, and phase parameters,
\begin{equation}
\begin{aligned}
    (\phi,\theta) =
    \left( \frac{\pi}{3}, \frac{5\pi}{12} \right),
    \qquad
    \kappa=1,
    \qquad
    \xi=-\frac{\pi}{2},
\end{aligned}
\end{equation}
with $\delta_0=0$ and $t_0=57.1~{\rm \mu s}$.

We then consider two representative chirping rates characterized by their ratio to the dissipation scale of the most sensitive mode ($a=3$ in the $\mathrm{TE}_{111+}$ family, with $Q_3^L=1.64\times10^3$),
\begin{itemize}
    \item Slow chirp: $h_0 = 4.9\times 10^{-19}$, $\alpha/\gamma_3^2 = 0.5$,
      \item Fast chirp: $h_0 = 1.5\times 10^{-18}$, $\alpha/\gamma_3^2 = 5$.
\end{itemize}
This mode provides the dominant response for the benchmark waveform, as shown in Fig.~\ref{fig:9cell_freqskymap}. The fast (slow) chirp sweeps across the full set of 18 cavity modes within approximately $37~{\rm \mu s}$ ($368~{\rm \mu s}$), over a bandwidth of $\sim 0.25~{\rm GHz}$. Both cases are normalized to yield a similar total signal-to-noise ratio, $\mathrm{SNR}=10$, when combining all cavity modes, as discussed in the next section.

The benchmark waveform can be interpreted as arising from a circular compact binary system observed face-on, with orbital inclination and polarization angles set to zero in the cavity frame. In this mapping, the waveform parameters $h_0$ and $\alpha$ are determined by the chirp mass $\mathcal{M}_c$ and luminosity distance $r$. For a circular inspiral~\cite{Maggiore:2018sht},
\begin{align}
    h_0
    &=
    \frac{4}{r}
    \left(G\mathcal{M}_c\right)^{5/3}
    \left(\frac{\omega_g}{2}\right)^{2/3},
    \\
    \alpha
    &=
    \frac{12}{5}\,2^{1/3}
    \left(G\mathcal{M}_c\right)^{5/3}
    \omega_g^{11/3}.
\end{align}
This mapping translates the benchmark parameters into representative compact binary configurations with
a chirp mass $\mathcal{M}_c \sim \mathcal{O}(10^{-4})~M_\oplus$, where $M_\oplus$ denotes the Earth mass, and a distance $r \sim \mathcal{O}(10^{-2})~{\rm AU}$.

The relatively small source distance can in principle be constrained by solar-system astrometric observations of long-term orbital precession. Current ephemeris-based limits can reach $M_{\rm exotic} \lesssim 10^{-6} M_\oplus$~\cite{Berlin:2026che}. However, such constraints apply primarily to static or bound configurations and do not directly constrain transient chirping sources. For unbound objects in the Galactic halo, the typical virial velocity ($\sim 220~{\rm km/s}$) exceeds the Solar System escape velocity ($\sim 30~{\rm km/s}$), making bound configurations unlikely.

\subsection{Waveforms in the Cavity}

The cavity response to such a signal is governed by the mode equation derived in Eq.~(\ref{eq:eommerged}). Here we focus on how the chirping drive generates a time-dependent waveform inside the cavity and sequentially excites different cavity modes.

\begin{figure}[h]
    \centering
    \includegraphics[width=\linewidth]{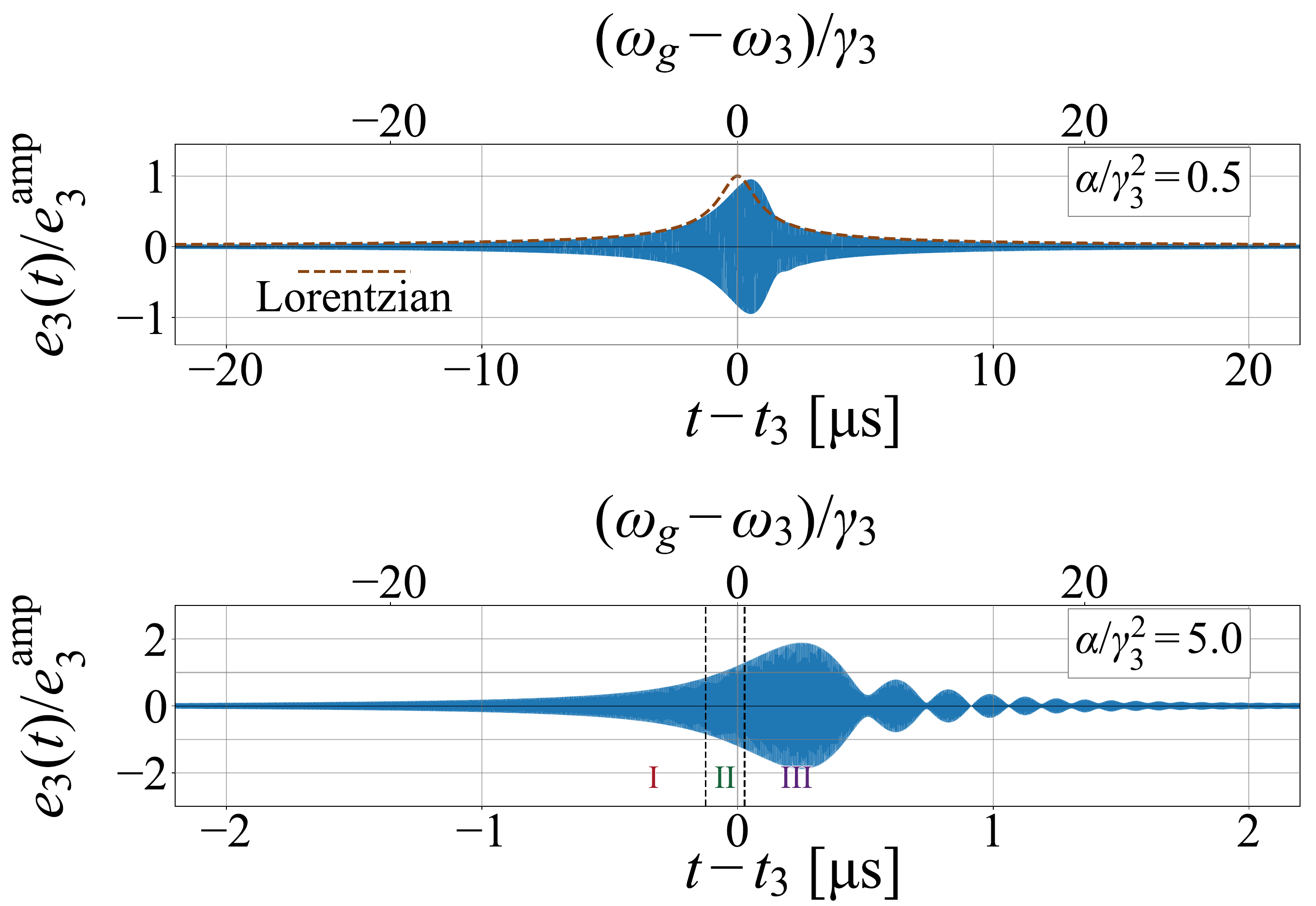}
    \caption{
    Time evolution of the dominant cavity mode, $a=3$ ($\mathrm{TE}_{111+}$), characterized by the mode amplitude $e_3(t)$, for the slow (top) and fast (bottom) chirp benchmark GW waveforms. The amplitudes are normalized by the corresponding approximate maximum values, $e_3^{\rm amp}$, defined in Eqs.~(\ref{eq:MaximumAmplitudeslow}) and (\ref{eq:MaximumAmplitude}), respectively.
    The time axis is shifted by $t-t_3$, where $t_3$ denotes the moment when the GW frequency $\omega_g = |\vec{k}|$ crosses the cavity resonance $\omega_3$. For the slow chirp, the dashed curve shows the Lorentzian response for comparison, while for the fast chirp, the three stages of the cavity response discussed in the text are indicated.
    }
    \label{fig:volt_signal}
\end{figure}

Figure~\ref{fig:volt_signal} shows the time-domain response of the $a=3$ mode in the $\mathrm{TE}_{111+}$ cavity family, obtained from Eq.~(\ref{eq:eommerged}). The signal grows as the chirping GW sweeps through the mode resonance and subsequently decays on a timescale of order $1/\gamma_3$. The physical features of this behavior will be discussed in the following.

\subsection{Green Function Solution and Compact Analytic Form}

Substituting the chirping waveform in Eq.~(\ref{eq:hPBH}) into the cavity equations, it is convenient to first combine the two polarization components into a single effective source term,
\begin{equation}
    \sum_{A=+,\times}h_A\,\eta_A^a
    \equiv
    h_0\,\eta_{\rm eff}^a
    \exp\left[
    \mi\left(
    \varphi_a+\omega_1(t-t_0)+\frac{\alpha}{2}(t-t_0)^2
    \right)
    \right],
    \label{eq:EoMsource}
\end{equation}
where the effective overlap function and phase are given by
\begin{align}
    \eta^a_{\rm eff}
    &\equiv
    \sqrt{\frac{
    |\eta_+^a|^2
    +
    \kappa^2|\eta_\times^a|^2
    +
    2\kappa|\eta_+^a\eta_\times^a|
    \cos\left(
    \Omega_\times^a-\Omega_+^a+\xi
    \right)}{1+\kappa^2}
    },
    \\
    \varphi_a
    &\equiv
    \delta_0+\Omega_+^a
    +
    \arctan\left[
    \frac{
    \kappa|\eta_\times^a|
    \sin(\Omega_\times^a-\Omega_+^a+\xi)
    }{
    |\eta_+^a|
    +
    \kappa|\eta_\times^a|
    \cos(\Omega_\times^a-\Omega_+^a+\xi)
    }
    \right],
    \label{eq:varphia}
\end{align}
with $\Omega^a_{+/\times}\equiv {\rm arg}[\eta^a_{+/\times}]$.

The mode equation in Eq.~(\ref{eq:eommerged}) can then be solved using the Green function of a damped harmonic oscillator, yielding
\begin{equation}
    e_a(t)
    =
    -\mi\omega_a^3B_0V^{5/6}
    \int_{-\infty}^{t}
    G_a(t-\tau)
    \left(
    \sum_{A=+,\times}h_A(\tau)\eta_A^a
    \right)
    {\rm d}\tau,
    \label{eq:esolution}
\end{equation}
where the Green function is
\begin{equation}
    G_a(t)
    =
    \me^{-\gamma_a t}
    \frac{
    \sin\left(\sqrt{\omega_a^2-\gamma_a^2}\,t\right)
    }{
    \sqrt{\omega_a^2-\gamma_a^2}
    }
    H[t],
    \label{eq:GreenFunction}
\end{equation}
with $H[t]$ the Heaviside step function.

Inserting Eq.~(\ref{eq:EoMsource}) into Eq.~(\ref{eq:esolution}) and defining $t' \equiv t - t_0$, the integral can be evaluated in closed form,
\begin{equation}
\begin{split}
     e_a(t)
     = &
     -\frac{\omega_a^2B_0V^{5/6}}{2}
     h_0\eta_{\rm eff}^a \me^{\mi\varphi_a}\\
     &\times
     \left[
     \me^{-\lambda_-^a t'}I_-^a(t')
     -
     \me^{-\lambda_+^a t'}I_+^a(t')
     \right],
     \label{eq:GeneralSolution}
     \end{split}
\end{equation}
where
\begin{equation}
    I_{\pm}^a(t')
    \equiv
    \sqrt{\frac{\pi}{2\alpha}}
    \me^{
    \mi\left(
    \frac{\pi}{4}
    +
    \frac{(b_\pm^a)^2}{4p}
    \right)}
    \left(
    {\rm erf}\left[
    \sqrt{p}
    t'-\frac{b_\pm^a}{2\sqrt{p}}
    \right]
    +1
    \right),
\end{equation}
with $\lambda_\pm^a\equiv
    \gamma_a
    \pm
    \sqrt{\omega_a^2-\gamma_a^2}$, 
    $b_\pm^a
    \equiv
    \gamma_a
    +
    \mi\left(
    \omega_1
    \pm
    \sqrt{\omega_a^2-\gamma_a^2}
    \right)$, and
    $p\equiv-\frac{\mi\alpha}{2}$.

\subsection{Approximate Time-Domain Behavior}

Although the exact solution in Eq.~(\ref{eq:GeneralSolution}) is useful for numerical evaluation, its physical behavior can be understood more transparently from the competition among three characteristic scales: the instantaneous detuning $|\omega_g(t)-\omega_a|$, the cavity damping rate $\gamma_a$, and the chirp rate $\alpha^{1/2}$. As the GW frequency crosses the cavity resonance, the response naturally separates into three regimes: off-resonant tracking, resonant growth, and post-resonance decay.

It is useful to define the instantaneous saturation amplitude, i.e. the steady-state response that would be obtained if the GW frequency were frozen at its instantaneous value,
\begin{equation}
    e_a^{\rm sat}(t)
    \equiv
    \left|
    \frac{
    \omega_a^3 B_0 V^{5/6}h_0\eta_{\rm eff}^a
    }{
    \omega_a^2-\omega_g(t)^2-2\mi\omega_g(t)\gamma_a
    }
    \right|.
    \label{eq:esat}
\end{equation}
The actual signal follows this moving saturation envelope but develops a finite lag when the chirp rate becomes comparable to or larger than the cavity response rate.

\paragraph{Slow chirp}
We first consider the regime of sufficiently slow frequency evolution,
\begin{equation}
    \alpha\leq 2\gamma_a^2,
\end{equation}
in which the cavity mode remains approximately adiabatically saturated throughout the resonance crossing. In this regime, the response is well approximated by
\begin{equation}
    |e_a(t)|
    \simeq
    e_a^{\rm sat}(t),
\end{equation}
and no distinct resonant amplification stage develops.

As shown in the upper panel of Fig.~\ref{fig:volt_signal}, the envelope closely follows the Lorentzian response of a damped harmonic oscillator, and this agreement becomes increasingly accurate for smaller values of $\alpha$. We normalize the evolution by the maximum amplitude of the Lorentzian response,
\begin{equation}
e_a^{\rm amp}
\equiv
\omega_a B_0 V^{5/6} h_0 \eta_{\rm eff}^a Q_a^L .
\label{eq:MaximumAmplitudeslow}
\end{equation}

\paragraph{Fast chirp}
We next consider the fast chirp regime,
\begin{equation}
    \alpha \geq 2\gamma_a^2,
    \label{eq:intermediatealpha}
\end{equation}
in which the frequency sweep is sufficiently rapid that the signal does not have enough time to reach the saturation value given in Eq.~(\ref{eq:esat}). Note that in deriving the approximations below, we further assume the adiabatic condition $\alpha/\omega_a < \gamma_a$, ensuring that the GW remains within the cavity bandwidth for several oscillation cycles so that an envelope description remains valid.

\noindent
$\bullet$ \emph{Stage I: off-resonant tracking:}  
At early times, when $|\omega_g(t)-\omega_a| \gg \sqrt{\alpha/2\pi}$, the GW drive is far off resonance and the saturation amplitude is small. In this regime the cavity mode remains effectively saturated, and the signal amplitude closely follows the instantaneous steady-state value,
\begin{equation}
    |e_a(t)|^{\rm (I)}
    \simeq
    e_a^{\rm sat}(t).
\end{equation}
Because the Lorentzian envelope in Eq.~(\ref{eq:esat}) grows only slowly far from resonance, the signal remains negligible during this stage.

As the GW frequency approaches $\omega_a$, the saturation amplitude increases rapidly. Stage~I ends at the time $t_a^s$ when the cavity can no longer adiabatically track this growth. A useful estimate for $t_a^s$ is obtained by equating the rate of change of the saturation amplitude to the intrinsic growth rate of the cavity mode, 
\begin{equation}
    \left.
    \frac{{\rm d}|e_a|}{{\rm d}t}
    \right|_{t=t_a^s}
    \sim
    \alpha
    \left.
    \frac{{\rm d}e_a^{\rm sat}}{{\rm d}\omega_g}
    \right|_{t=t_a^s}
    \sim
    \omega_a^2B_0V^{5/6}h_0\eta_{\rm eff}^a .
\end{equation}

\noindent
$\bullet$ \emph{Stage II: resonant growth:}  
For $t > t_a^s$, the instantaneous saturation amplitude exceeds the actual signal amplitude, and the cavity mode undergoes resonant energy accumulation. 
During this stage, the signal grows approximately as an effectively monochromatic resonant response, with the growth controlled by the cavity dissipation rate $\gamma_a$,
\begin{equation}
    |e_a(t)|^{\rm (II)}
    \simeq
    \left(
    1-\me^{-\gamma_a(t-t_a^s)}
    \right)
    \left(
    \omega_a B_0V^{5/6}h_0\eta_{\rm eff}^a Q_a^L
    \right).
\end{equation}
This growth persists only while the GW frequency remains sufficiently close to resonance. Once $\omega_g(t)$ exceeds $\omega_a$, the saturation amplitude $e_a^{\rm sat}(t)$ decreases rapidly. Stage~II ends at time $t_a^e$, defined by the condition
\begin{equation}
    e_a^{\rm sat}(t_a^e)
    \sim
    |e_a(t_a^e)|^{\rm (II)}.
\end{equation}

The duration of the resonant growth window is primarily controlled by the frequency sweep rate and can be estimated as
\begin{equation}
    t_a^e-t_a^s
    \sim
    \frac{1}{\sqrt{\alpha}}.
\end{equation}
This timescale is parametrically longer than the naive estimate based on the interval during which the GW frequency remains within the cavity bandwidth, $\sim \gamma_a/\alpha$. This indicates that significant signal growth can occur even when the GW frequency lies outside the nominal resonance bandwidth, as the cavity mode has not yet reached its saturation limit.

\noindent
$\bullet$ 
 \emph{Stage III: post-resonance decay:}  
For $t > t_a^e$, the driving frequency has moved off resonance and the cavity mode undergoes free decay with damping rate $\gamma_a$, accompanied by weak beat oscillations. The signal amplitude in this stage is approximately
\begin{equation}
\begin{split}
    & |e_a(t)|^{\rm (III)}\\
    \simeq &
    \left(
    \omega_a B_0V^{5/6}h_0\eta_{\rm eff}^a Q_a^L
    \right)
    \left(
    1-\me^{-\gamma_a(t_a^e-t_a^s)}
    \right)
    \me^{-\gamma_a(t-t_a^e)} .
    \end{split}
\end{equation}

Since the signal growth during Stage~I is negligible, the maximum oscillation amplitude is well approximated by the value reached at the end of Stage~II,
\begin{equation}
    e_a^{\rm amp}
    \equiv
    |e_a(t_a^e)|
    \simeq
    \left(
    1-e^{-\gamma_a/\sqrt{\alpha}}
    \right)
    h_0B_0\omega_a V^{5/6}Q_a^L\eta_{\rm eff}^a .
    \label{eq:MaximumAmplitude}
\end{equation}
The subsequent decay occurs on the timescale $1/\gamma_a$. 

Averaging over fast oscillations, the time-integrated signal power is approximately
\begin{equation}
\begin{aligned}
   & \int_{-\infty}^{\infty} \left(\frac{e_a(t)}{e_a^{\rm amp}}
    \right)^2\,{\rm d}t\\
    \approx &
    \frac{1}{2}\int_{t_a^s}^{t_a^e}
    \left(
    1-\me^{-\gamma_a(t-t_a^s)}
    \right)^2
    {\rm d}t +
    \frac{1}{2}\int_{t_a^e}^{\infty}
    \me^{-2\gamma_a(t-t_a^e)}
    {\rm d}t
    \\
    \simeq &
    \frac{1}{4\gamma_a}
    \left(
    1+
    \mathcal{O}
    \left(
    \frac{\gamma_a}{\sqrt{\alpha}}
    \right)^3
    \right).
\end{aligned}
\end{equation}

As illustrated in the lower panel of Fig.~\ref{fig:volt_signal}, the fast chirp regime exhibits the full three-stage evolution described above. For larger values of $\alpha$, the growth phase becomes increasingly short, while the subsequent decay remains universally governed by the cavity dissipation rate $\gamma_a$.


\section{Dissecting Gravitational Wave Strain from Multi-Cell Responses}

\begin{figure*}[h]
    \centering
    \includegraphics[width=\textwidth]{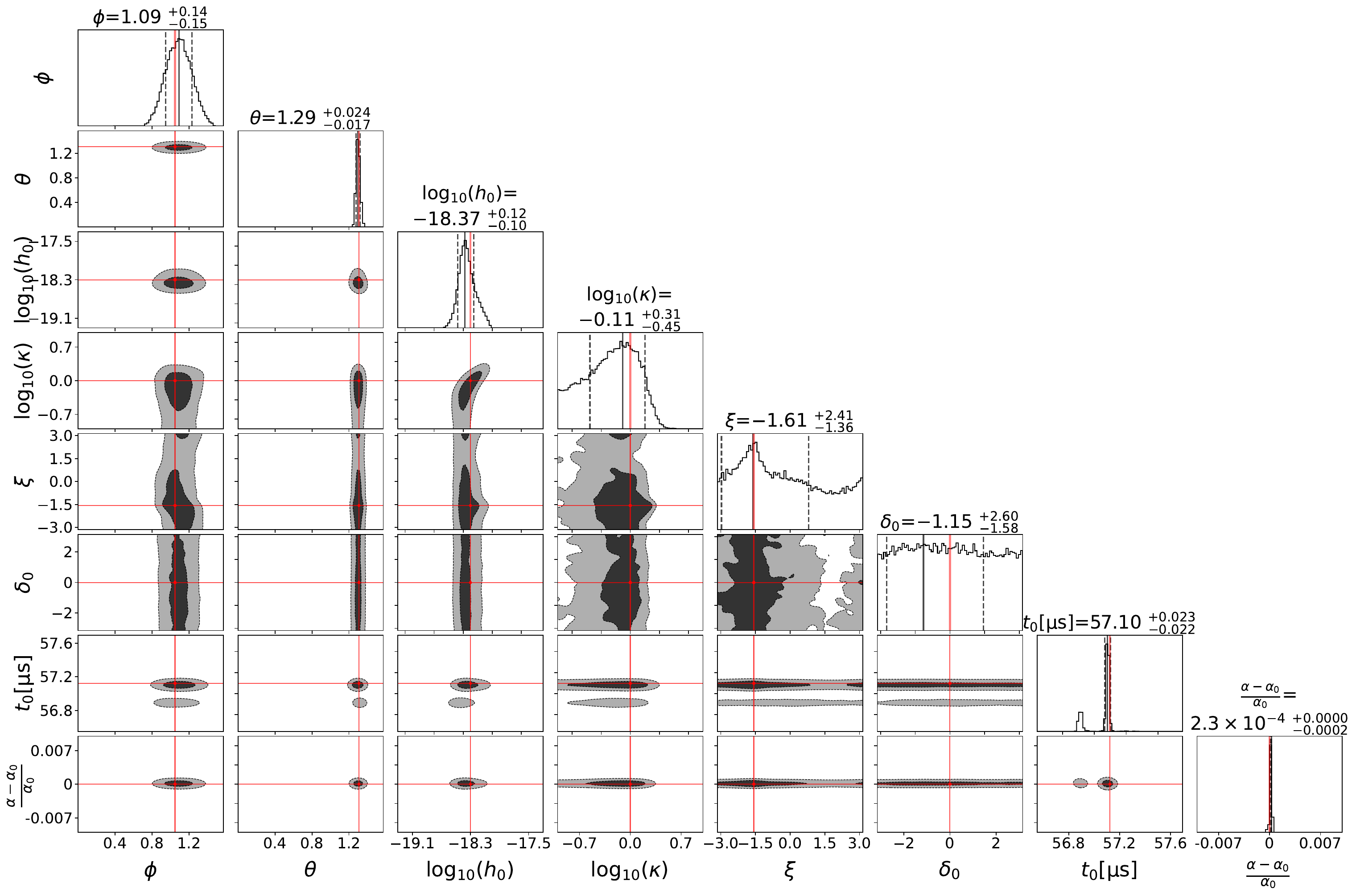}
    \caption{Posterior distributions of the GW waveform parameters $\vec{\Theta} = (\phi, \theta, h_0, \kappa, \xi, \delta_0, t_0, \alpha)$ for the slow chirp waveform. Here $(\phi,\theta)$ denote the GW propagation direction, $h_0$ is the GW strain amplitude, $\kappa$ and $\xi$ are the two polarizations’ ratio and relative phase, $t_0$ is the time at which the GW frequency reaches the first cavity resonance with phase $\delta_0$, and $\alpha$ is the frequency drift rate. The red solid lines indicate the true parameter values. On the diagonal panels, black solid lines mark the posterior modes, while dashed lines indicate the $1\sigma$ ($68\%$) credible intervals; the inferred values and corresponding uncertainties are listed at the top of each panel. In the off-diagonal panels, the shaded regions indicate the $39.3\%$ and $86.5\%$ credible regions of the marginalized two-dimensional posterior distributions, respectively. Uniform priors are assumed within the figure range for each parameter.} 
    \label{fig:MCMC}
\end{figure*}

\begin{figure*}[h]
    \centering
    \includegraphics[width=\textwidth]{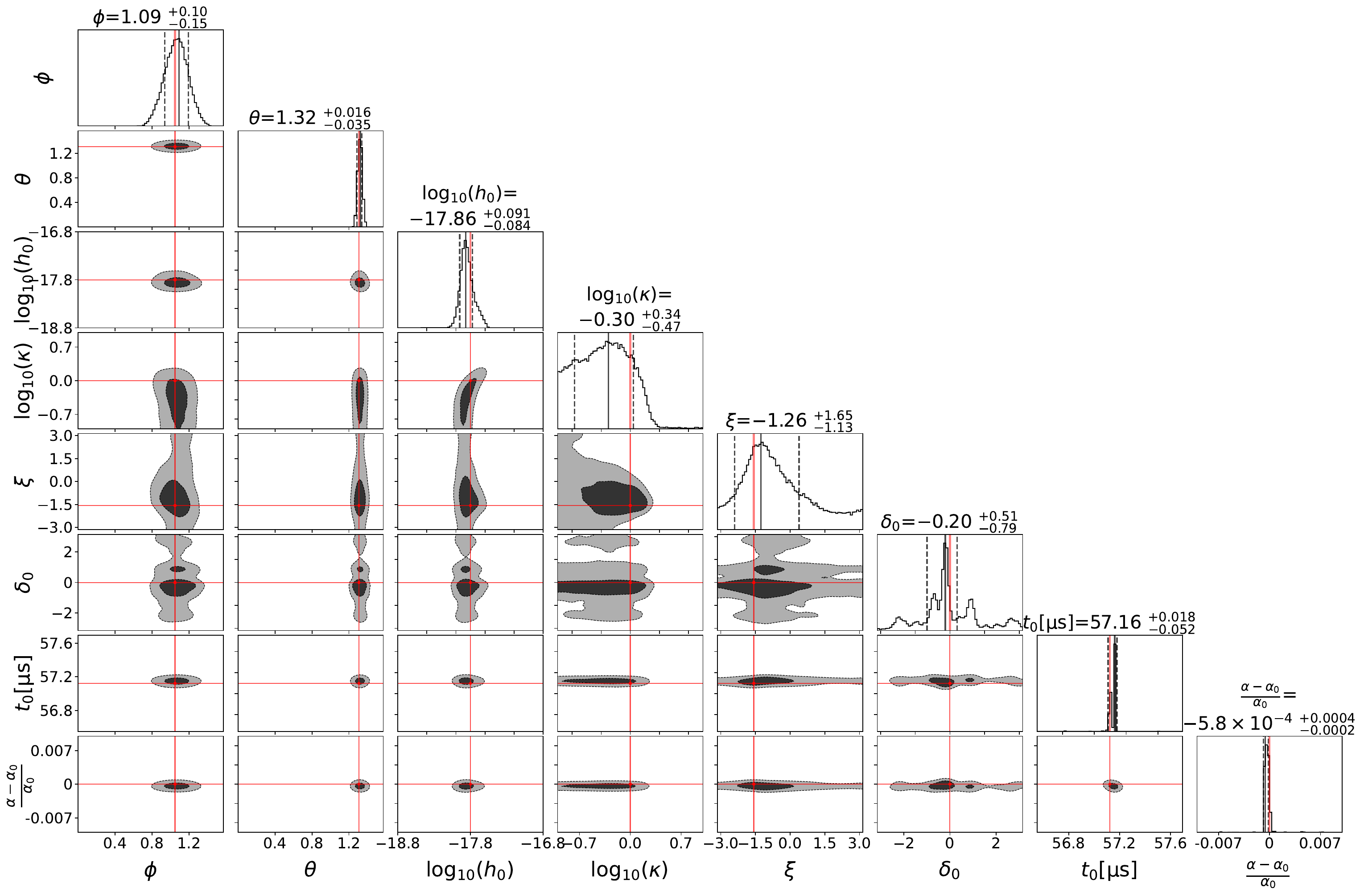}
    \caption{Same as Fig.~\ref{fig:MCMC} for the fast chirp waveform.}
    \label{fig:MCMC_fast_appendix}
\end{figure*}

We generate a mock data stream $d_a^i = e_a(t_i) + n_a(t_i)$ consisting of the benchmark signal plus Gaussian noise,
\begin{equation}
    n_a^i \sim \mathcal{N}(0,\sigma_a^2), 
    \qquad 
    \sigma_a^2 = \frac{2\,T_{\rm eff}}{\omega_a \Delta t},
\end{equation}
where $T_{\rm eff}$ is the effective noise temperature. The sampling interval is 
$\Delta t = t_{\rm tot}/N_t$, with $t_{\rm tot}$ the total measurement time and $N_t$ the number of samples; $\Delta t$ must be short enough to resolve the relevant frequencies.

We consider the $8$ HFGW parameters appearing in Eq.~(\ref{eq:hPBH}), $\vec{\Theta} = (\phi,\theta,h_0,\kappa,\xi,\delta_0,t_0,\alpha)$,
and construct the likelihood for a given parameter set as
\begin{equation}
\begin{aligned}
\mathcal{L}(\vec{\Theta}\,|\,d_a^i)
= \prod_{a} 
\left( 2\pi\sigma_a^{2} \right)^{-\frac{N_t}{2}}
\exp\!\left[
-\sum_{i}\frac{\left( d_a^i - e_a^{\Theta}(t_i) \right)^2}{2\,\sigma_a^2}
\right],
\end{aligned}
\end{equation}
where the product runs over all $18$ cavity modes. The Fisher information matrix is then obtained by evaluating
\begin{equation}
    F_{jk}
    =
    \left.
    \Big\langle
    \frac{\partial^2 [-\ln\mathcal{L}]}{\partial\Theta_j\,\partial\Theta_k}
    \Big\rangle
    \right|_{\vec{\Theta}=\vec{\Theta}_{\rm max}},
\end{equation}
at the maximum-likelihood point $\vec{\Theta}_{\rm max}$.

The approximate signal-to-noise ratio (SNR) follows from
\begin{equation}
\begin{aligned}
{\rm SNR}^2
&= F_{h_0 h_0} \, h_0^2 
= \sum_{a}\sum_{i}
\left( \frac{e_a(t_i)}{\sigma_a} \right)^2 \\
&\approx 
\frac{h_0^2 B_0^2 V^{5/3}}{2\alpha T_{\rm eff}}
\sum_{a} Q_a^L \left(\omega_a^2 \eta_{\rm eff}^a \right)^2.
\end{aligned}
\label{snreq}
\end{equation}
Thus the sensitivity scales as $h_0 \sim N^{-1/2}$, effectively behaving as an $N$-detector network.  

We adopt the benchmark detector parameters
$B_0 = 4~\mathrm{T}$,
$V = 25~\mathrm{L}$,
and $T_{\rm eff} = 100~\mathrm{mK}$, considering the $18$ cavity modes listed in Table~\ref{tab:fqletasnr}. 
For both benchmark GW waveforms, we compute the mode-resolved contributions to the signal-to-noise ratio $\mathrm{SNR}_a^2$, together with the corresponding effective overlap functions $\eta_{\mathrm{eff}}^a$, as listed in Table~\ref{tab:fqletasnr}.
In both cases, the total combined sensitivity reaches $\mathrm{SNR}\simeq 10$.

To reconstruct the full signal, at least five cavity modes must be clearly resolved. The relative signal strengths provide access to five waveform parameters, including the overall strain amplitude, propagation direction, polarization ratio, and polarization phase. Time-domain information further determines three additional parameters: the frequency drift rate, the initial time, and the overall phase, yielding a total of eight independent degrees of freedom of the signal. In our mock data analysis, seven cavity modes achieve $\mathrm{SNR}_a > 1$, providing sufficient information for a stable reconstruction of all waveform parameters.

Fig.~\ref{fig:MCMC} and Fig.~\ref{fig:MCMC_fast_appendix} show the corresponding Markov chain Monte Carlo (MCMC) posterior distributions for the slow and fast chirp waveforms, respectively. In both cases, as expected, all $8$ parameters are successfully reconstructed.

The parameters associated with the signal strength, $(\phi,\theta,h_0,\kappa,\xi)$, are well reconstructed, in agreement with simple Fisher-matrix estimates, with uncertainties scaling as $\Delta\Theta_j \sim F_{jj}^{-1/2} \propto \mathrm{SNR}^{-1}$. In particular, the directional parameter $\theta$ exhibits a much smaller relative uncertainty than $\phi$, reflecting the sharply peaked angular response in the $\theta$ direction.

In contrast, the reconstruction of the initial phase $\delta_0$ exhibits a strong dependence on the chirp regime. This is primarily due to the fact that both $\delta_0$ and $t_0$ are defined at the time when the GW frequency first matches the lowest cavity mode, which has only marginal sensitivity, with ${\rm SNR}_1 \simeq 0.66$. As a result, the uncertainty in $\delta_0$ is largely inherited from the propagation of phase information from the most sensitive mode ($a=3$) back to the lowest mode, and is therefore sensitive to the uncertainty in the chirp rate $\alpha$. This effect is more pronounced in the slow-chirping case, where the longer time interval between successive mode excitations leads to enhanced propagation of phase uncertainty. The posterior distribution of $t_0$ further develops secondary peak structures, arising from the periodic dependence of the relative phases.

The reconstructed signal parameters can be mapped onto a complete set of binary parameters with the same number of degrees of freedom. Notably, the $\mathcal{M}_c$–$r$ degeneracy can be broken using frequency-drift information.

\section{Discussion}

HFGWs remain a largely unexplored frontier in gravitational physics, carrying potential signatures of phenomena beyond the Standard Model. EM resonant cavities, widely used in axion searches, provide a powerful platform for HFGW detection owing to their analogous microscopic coupling to EM fields. As tensorial excitations, GWs encode rich information about their sources and production mechanisms in their waveform, polarization, and propagation direction. In this work, we exploit multiple nearly degenerate modes within a single cavity to reconstruct the waveform properties of incoming HFGWs. Achieving this reconstruction requires at least $5$ sufficiently loud modes, which motivates the use of multi-cell cavities to ensure an adequate number of detectable modes. The sensitivity scaling with the number of modes is analogous to that of a network comprising the same number of independent detectors, making a single cavity significantly more capable of probing astrophysically plausible binary sources.

Our method of dissecting GW signals through comparisons among modes with distinct antenna patterns demonstrates that a single cavity can already function as an effective detector array, without the need for phase synchronization. The multi-cell configuration can be naturally generalized to a network of multiple detectors~\cite{Schmieden:2023fzn,Schneemann:2024qli}. Combining internal modal diversity with spatially separated cavities, such as in a long-baseline configuration~\cite{Jiang:2023jhl}, would enable even richer searches and reconstructions of HFGWs.

While we have focused on HFGWs from inspiraling PBH binaries with relatively simple waveforms, more complex signals, including merger and ringdown phases~\cite{Maggiore:2018sht}, can be naturally incorporated within our framework. In such cases, upgrading the cavity to a readout system capable of simultaneous resonant and broadband operation could significantly enhance sensitivity~\cite{Chen:2021bgy,Wurtz:2021cnm,Jiang:2022vpm,Chen:2023ryb}. It would also be valuable to extend our analysis from time-series voltage measurements to photon-counting detection ~\cite{Dixit:2020ymh,Agrawal:2023umy,Zheng:2025qgv,Kang:2025kaf}, which can improve the signal-to-noise ratio while sacrificing full time-domain information.

\begin{acknowledgments}

We are grateful to Asher Berlin, Hanyu Cheng, Tom Krokotsch, Zhen Liu, Nick Rodd, Bo Wang, Daixu Yang, and Yanjie Zeng for useful discussions. 
The Center of Gravity is a Center of Excellence funded by the Danish National Research Foundation under grant No. 184.
Y.C. is supported by the Shanghai Qiguang Natural Science Development Foundation.
J.S. is supported by the Peking University under startup Grant No. 7101302974, and the National Natural Science Foundation of China under Grants No. 12025507, No. 12450006.
This publication is part of the grant PID2023-146686NB-C31 funded by MICIU/AEI/10.13039/501100011033/ and by FEDER, UE.
IFAE is partially funded by the CERCA program of the Generalitat de Catalunya.
This work is supported by ERC grant (GravNet, ERC-2024-SyG 101167211, DOI: 10.3030/101167211). Funded by the European Union. Views and opinions expressed are however those of the author(s) only and do not necessarily reflect those of the European Union or the European Research Council Executive Agency. Neither the European Union nor the granting authority can be held responsible for them.
D.B. acknowledges the support from the European Research Area (ERA) via the UNDARK project of the Widening participation and spreading excellence programme (project number 101159929).

\end{acknowledgments}

\providecommand{\href}[2]{#2}\begingroup\raggedright\endgroup

\end{document}